\documentclass[12pt,english]{article}
\usepackage[T1]{fontenc}
\usepackage[koi8-r]{inputenc}
\usepackage{geometry}
\usepackage{babel}
\usepackage{graphics}

\makeatletter

\providecommand{\LyX}{L\kern-.1667em\lower.25em\hbox{Y}\kern-.125emX\@}

\makeatother
\begin{document}

\title{The pair production of charmed mesons in the photon-photon interaction.}

\date{{}}

\author{A.V. Berezhnoy, V.V. Kiselev, A.K.Likhoded }

\maketitle
\begin{abstract}
In the framework of the constituent quark model the exclusive pair
production of the charmed mesons in the photon-photon interaction
is calculated. The comparison of these predictions with the ones of
the heavy quark effective theory is performed. It is shown, that the
light valence quark of D-meson plays the essential role not only in
the hadronization process, but also in the production of the heavy
c-quark. Moreover, it is shown, that because of the strong interaction
between the initial photon field and the light quark charge the similar
situation is kept even in the limit \( m_{Q}\rightarrow \infty  \)
, whence it is concluded, that the application of the heavy quark
effective theory in the case of the photon-photon interaction is not
valid because it does not take into account one of the dominant mechanisms
of the heavy meson production.
\end{abstract}

\section*{Introduction}

Recent years the big volume of work devoted to the experimental study
of the photon-photon interactions, in particular, to the study of
the photonic production of the charmed particles was carried out at
the accelerator LEP. In this connection, and also in connection with
prospects for construction of the largest \( e^{+}e^{-} \)-collider
TESLA (DESY) there is a necessity of the detailed discussion of the
photon-photon interaction mechanisms at large energies. In the given
work we shall consider some questions concerning the photonic production
of heavy quarks, namely, we shall discuss features of the exclusive
and the inclusive charmed meson pair production and estimate its contribution
into the total cross section of the charm production. In the frame
work of the constituent quark model (CQM) we shall show, that the
light valence quark inside D-meson plays the essential role not only
in the hardonization process, but also in the hard production of the
heavy quark. Because of strong interaction of the photonic field with
the light quark charge the similar situation is kept even in a limit
\( m_{Q}\rightarrow \infty  \).

It is worth to notice, that for the exclusive pair production of the
charmed meson in the \( e^{+}e^{-} \)-annihilation CQM and the heavy
quark effective theory (HQET) give the same prediction for \( m_{Q}\rightarrow \infty  \).
Nevertheless at the conventional values of masses of the light and
heavy quarks the results of CQM for the \( e^{+}e^{-} \)-interactions
strongly differ from the results of the HQET.

\section{The pair production of the charmed mesons in the \protect\( e^{+}e^{-}\protect \)-annihilation.}

We begin our article with the discussion of the heavy meson production
in the \( e^{+}e^{-} \)-annihilation, because it is the most simple
process, and because it is considered in detail in the literature. 

The work\cite{Rujula} presents one of the first attempts to formulate
the HQET for the case of the \( D \)-meson production in the \( e^{+}e^{-} \)-annihilation.
The neglect of spin-spin interaction between the heavy and light quarks,
which is suppressed as \( 1/m_{Q} \), has allowed to obtain the interesting
ratio for the pair production of \( D \) and \( D^{*} \)-meson near
the threshold (see the Appendix): \begin{equation}
\label{eeotn0}
\sigma _{D\bar{D}}:\sigma _{D\bar{D}^{*}+D^{*}\bar{D}}:\sigma _{D^{*}\bar{D}^{*}}=1:4:7
\end{equation}

More detailed analysis of the exclusive \( D\bar{D} \)-pair production
in the \( e^{+}e^{-} \)-interaction on the basis of the CQM, was
carried out in work \cite{Kiselev}. It is worth to remind, that CQM
is based on the assumptions, that in the partonic distributions of
the final meson there are the valence quark contributions, and that
both valence quarks are produced in the hard process followed by the
fusion of these quarks into the meson. 

The analytical expressions for the exclusive meson pair production
look as follows \cite{Kiselev}:\[
\sigma (e^{+}e^{-}\rightarrow (Q\bar{q})_{P}(\bar{Q}q)_{P})=\frac{\pi ^{3}\alpha _{s}^{2}(4m_{q}^{2})\alpha ^{2}_{em}}{3^{7}4m^{6}_{q}}\frac{m^{2}_{Q}}{M^{2}}f^{4}_{P}(1-v^{2})v^{3}\times \]

\begin{equation}
\label{eepp}
\times \left\{ 3e_{Q}\left( \frac{2m_{q}}{m_{Q}}-1+v^{2}\right) -3e_{q}\left[ 2-(1-v^{2})\frac{m_{q}}{m_{Q}}\right] \frac{m^{3}_{q}\alpha _{s}(m^{2}_{Q})}{m^{3}_{Q}\alpha _{s}(4m^{2}_{q})}\right\} ^{2},
\end{equation}

\begin{equation}
\label{eepv}
\sigma (e^{+}e^{-}\rightarrow (Q\bar{q})_{P}(\bar{Q}q)_{V})=\frac{\pi ^{3}\alpha _{s}^{2}(4m_{q}^{2})\alpha ^{2}_{em}}{3^{7}4m^{6}_{q}}\frac{m^{2}_{Q}}{M^{2}}f^{2}_{P}f_{V}^{2}(1-v^{2})v^{3}\left[ 3e_{Q}+3e_{q}\frac{m^{3}_{q}\alpha _{s}(m^{2}_{Q})}{m^{3}_{Q}\alpha _{s}(4m^{2}_{q})}\right] ^{2},
\end{equation}

\[
\sigma (e^{+}e^{-}\rightarrow (Q\bar{q})_{V}(\bar{Q}q)_{V})=\frac{\pi ^{3}\alpha _{s}^{2}(4m_{q}^{2})\alpha ^{2}_{em}}{3^{7}4m^{6}_{q}}f^{4}_{V}(1-v^{2})v^{3}\left[ 3e_{Q}-3e_{q}\frac{m^{3}_{q}\alpha _{s}(m^{2}_{Q})}{m^{3}_{Q}\alpha _{s}(4m^{2}_{q})}\right] ^{2}\times \]

\begin{equation}
\label{eevv}
\times \left[ 3(1-v^{2})+(1+v^{2})(1-a)^{2}+\frac{a^{2}}{2}(1-v^{2})(1-3v^{2})\right] ,
\end{equation}
where \( m_{q} \) and \( m_{Q} \) are masses of the light and heavy
quarks correspondingly; \( f_{P} \) and \( f_{V} \) are leptonic
coupling constants for the pseudoscalar and vector meson states correspondingly;
\( M=m_{q}+m_{Q} \); \( v=\sqrt{1-4M^{2}/s} \); and the parameter
\( a \) is expressed by the formula:

\[
a=\frac{m_{Q}}{M}\frac{1-\frac{e_{q}}{e_{Q}}\frac{m^{4}_{q}}{m^{4}_{Q}}\frac{\alpha _{s}(4m^{2}_{Q})}{\alpha _{s}(4m^{2}_{q})}}{1-\frac{e_{q}}{e_{Q}}\frac{m^{3}_{q}}{m^{3}_{Q}}\frac{\alpha _{s}(4m^{2}_{Q})}{\alpha _{s}(4m^{2}_{q})}}.\]
 At \( m_{Q}\rightarrow \infty  \) the ratio of the cross sections
on the threshold looks like: 

\begin{equation}
\label{eeotn}
\sigma _{PP}:\sigma _{PV}:\sigma _{VV}=1:4\frac{f^{2}_{V}}{f^{2}_{P}}:7\frac{f^{4}_{V}}{f^{4}_{P}}.
\end{equation}

Thus if one neglects the difference between the values of \( f_{V} \)
and \( f_{P} \), then(\ref{eeotn}) coincides with(\ref{eeotn0}).

The heavy quark effective theory predicts the ratio which at a threshold
also coincides with (\ref{eeotn0}):

\[
(1+h):\frac{s}{M^{2}}:3\left( 1+\frac{s}{3M}+h\right) ,\]
where the \( \alpha _{s} \) -correction has the following form: \[
h=-\frac{2\alpha _{s}}{3\pi }{\sqrt{1-\frac{4M^{2}}{s}}}\ln \left( \frac{s}{2M^{2}}-1+\frac{s}{2M^{2}}{\sqrt{1-\frac{4M^{2}}{s}}}\right) .\]

\begin{figure}
{\centering \resizebox*{0.8\columnwidth}{!}{\includegraphics{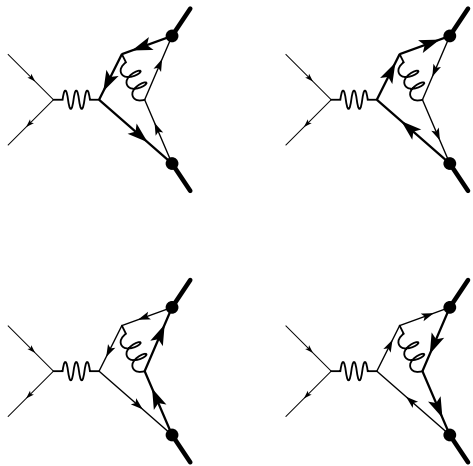}} \par}

\caption{The diagrams of the meson production in the \protect\( e^{+}e^{-}\protect \)-annihilation\label{eediagr}. }
\end{figure}

The four Feynmann diagrams of \( O(\alpha ^{2}_{s}\alpha ^{2}_{em}) \)
order contribute to the cross sections (\ref{eepp}), (\ref{eepv}),
(\ref{eevv}) (see Fig.\ref{eediagr}) . These diagrams can be separated
into two gauge invariant pairs. The first pair (two top diagrams of
Fig.\ref{eediagr}) corresponds to the interaction of the virtual
photon with the heavy quark and is proportional to the heavy quark
charge squared. The value of \( \alpha _{s} \) for this diagram pair
is calculated at the scale of \( 4m^{2}_{q} \), because the gluonic
propagator couples with the two light quarks. The second diagram pair
(the two bottom diagrams of the Fig. \ref{eediagr}) corresponds to
the interaction of virtual photon with the light quark and is proportional
to the light quark charge. For these diagrams \( \alpha _{s} \) is
calculated at the scale \( 4m^{2}_{Q} \). It is clear, that this
contribution is suppressed by the factor \( m^{2}_{q}/m^{2}_{Q} \)
due to the gluonic propagator coupled with the pair of heavy quarks.

One can see from (\ref{eepp}), (\ref{eepv}), (\ref{eevv}) , that
at \( m_{Q}\rightarrow \infty  \) it is possible to neglect the interaction
of the light quark with the initial virtual photon. It is necessary
to note, that the contribution of that interaction is small already
for the masses of constituents of the \( D \)-meson (for present
calculations the following values of masses were chosen: \( m_{c}=1.8 \)
~GeV and \( m_{q}=0.2 \)~GeV). However, the spin-spin interaction
at these values of masses plays the essential role and brakes the
ratio (\ref{eeotn0}): \[
\sigma _{D\bar{D}}:\sigma _{D\bar{D}^{*}+D^{*}\bar{D}}:\sigma _{D^{*}\bar{D}^{*}}\approx 1:8:14.\]

So, the \( D\bar{D} \)-pair production in the \( e^{+}e^{-} \)-annihilation
occurs as follows: a pair of heavy quarks is produced in a hard process,
and then these quarks are hadronized by means of interaction with
the light quark sea. It is worth to remind, that in the inclusive
production of \( D \)-mesons in the \( e^{+}e^{-} \)-annihilation,
where the fragmentation mechanism can be applied\cite{frag} , the
\( c \)-quarks are hadronized by the same manner. Thus, the role
of light quark in the hard process of the \( c \)-quark production
in the \( e^{+}e^{-} \)-annihilation is insignificant. On the contrary,
in the photon-photon production the light quark role is essential
as will be shown bellow.

\section{The exclusive \protect\( D\bar{D}\protect \)-pair production in
the photon-photon interaction}

\begin{figure}
{\centering \resizebox*{0.8\columnwidth}{!}{\includegraphics{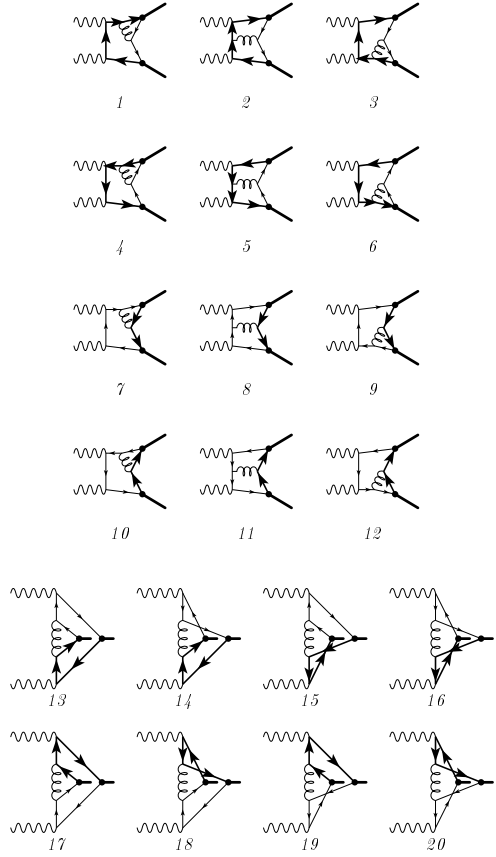}} \par}

\caption{The diagrams of the photonic production of two heavy mesons. \label{diagrams}}
\end{figure}

In the framework of the CQM the photonic exclusive production of two
charmed mesons is described by twenty Feynmann tree level diagrams
(see the Fig. \ref{diagrams}), which can be separated into three
gauge invariant groups. The first of them corresponds to a case when
the heavy quark radiate the gluon which splits into a pair of light
quarks (diagrams 1-6 of the Fig. \ref{diagrams}). The contribution
of this group is proportional to the square of the heavy quark charge.
In the second group of diagrams (the diagram 7-12), on the contrary,
the heavy quark production occurs by means of the splitting of the
gluon which is radiated by the light quark. These diagrams are proportional
to the square of the light quark charge and suppressed by the factor
\( m_{q}^{2}/m^{2}_{Q} \) due to the gluonic propagator which is
proportional to \( 1/m^{2}_{Q} \). 

In the third group of diagrams (the diagram of 13-20 Fig. \ref{diagrams})
each pair of quarks is connected to its {}``own{}`` \( \gamma  \)-quantum.
Our calculations show, that the interaction of the light quark with
the initial photon field described by these diagrams is not suppressed
and plays the essential role in the exclusive pair production of the
charmed meson in the photon-photon interaction. This interaction leads
to the discrepancy between the predictions of HQET and CQM, which
is not negligible even at \( m_{Q}\rightarrow \infty  \). Moreover,
the interaction of the light constituent quark with the initial photon
is so large, that HQET can not be applied in this case.

Nevertheless, if one puts the light quark charge equal to zero in
our calculations (this case corresponds to the rejection of the diagrams
of the second and third groups, i.e. 7-20 of Fig. \ref{diagrams})
then the conformity between CQM and HQET is restored. Thus, at zero
charge of the light quark and \( m_{Q}\rightarrow \infty  \) we have
received on the threshold the following ratio:\begin{equation}
\label{ratio3}
\sigma _{PP}:\sigma _{PV}:\sigma _{VV}=1:0:3,
\end{equation}
 predicted in the work\cite{Kartvelishvili}, where the approach \cite{Rujula}
is applied to the case of photonic production (see Appendix). If the
light quark charge is differ from zero then its interaction with the
photonic field becomes essential for the \( D\bar{D} \)-pair production.
Moreover, our calculation show, that \( m_{Q}\rightarrow \infty  \)
and \( 2m_{Q}/s\rightarrow 1 \) lead to \( \sigma _{VV}/\sigma _{PP}\rightarrow \infty . \)
It means that the mechanism of the exclusive production based on the
production of the heavy quark pair followed by their hadronization
appears not valid. As it is shown in the work \cite{BKL}, such approach
is not valid in the case of the inclusive production too.

\begin{figure}
\hfill\parbox{12.5cm}{\( \sigma  \), nb}

{\centering \resizebox*{0.5\textwidth}{!}{\includegraphics{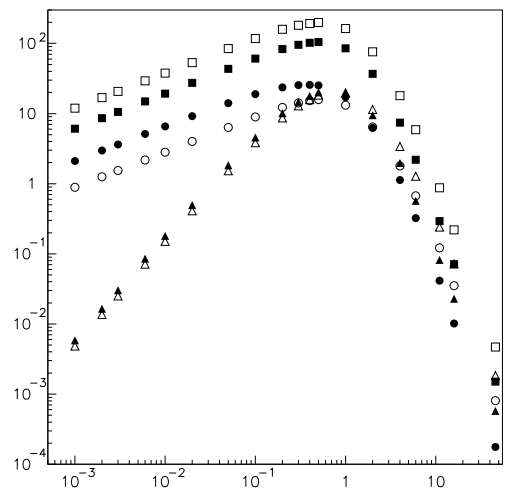}} \par}

{\centering \parbox{8cm}{\hfill \( \Delta  \), GeV}\par}

\caption{The exclusive cross section of the photonic pair production of the
charged charmed mesons \protect\( \sigma _{PP}\protect \) (the open
circles), \protect\( \sigma _{PV}\protect \) (the open triangles),
\protect\( \sigma _{VV}\protect \) (the open squares) as a function
of \protect\( \Delta ={\sqrt{s}}-{\sqrt{s_{th}}}\protect \). The
black markers show the same dependences at a zero charge of the light
quark\label{DDcharged}. }
\end{figure}

In the Fig. ~\ref{DDcharged} we present the exclusive cross sections
of the photonic pair production of the charmed mesons \( \sigma _{PP} \)
(open circles), \( \sigma _{PV} \) (open triangles), \( \sigma _{VV} \)
(open squares) calculated by us in the framework of the CQM as a function
of \( \Delta ={\sqrt{s}}-{\sqrt{s_{th}}} \) with the following values
of parameters:\[
m_{c}=1.5\rm \; GeV;\]
\[
m_{q}=0.3\rm \; GeV;\]
\[
\alpha _{s}=0.3.\]

As well as in the case of production in the \( e^{+}e^{-} \)-annihilation,
the cross sections \( \sigma _{PP} \), \( \sigma _{PV} \) and \( \sigma _{VV} \)
are proportional to \( f_{P}^{4} \), \( f_{P}^{2}f_{V}^{2} \) and
\( f_{V}^{4} \), accordingly.

The values of \( f_{P} \) and \( f_{V} \) estimated in the frame
work of the sum rules, as well as in the frame work of the lattice
calculations are about \( f_{P}\sim f_{V}\sim 200-300\rm \; MeV \).
However, the absolute normalization of the cross sections in our model
depends not only on these constants, but also on the light quark mass.
The variation of the light quark mass can result in the change of
the cross section normalization.

On the other hand, we can normalize the values of \( f_{P} \) and
\( f_{V} \) on the base of the fragmentation probability \( c\rightarrow D^{*} \)
in the \( e^{+}e^{-} \)-interaction at large energies (\( W(c\rightarrow D^{*})=0.22 \)).
Thus the values of \( m_{q} \) and \( f_{P,V} \) appear correlated
to each other, that allows to minimize the cross section uncertainty
within 2-2.5. However, with such normalization the constant \( f_{P,V} \)
appears overestimated at all reasonable values of the light quark
mass. We believe, that this overestimate of the constant \( f_{P,V} \)
is due to the account for the contribution of the excited state decays,
for example, of \( D^{**}\rightarrow D^{*}+X \). In the exclusive
production such situation should not be observed. Thus, the question
of the absolute normalization for the exclusive channels remains open
and the values in Fig.\ref{DDcharged} should be considered as the
top limit of the cross section. For all conclusions connected to the
relative yield of \( D \)-mesons the discussed circumstances are
not essential.

Let us notice, that near the threshold \( \Delta \approx |k|^{2}/M \),
where \( |k| \) is the module of three-momentum of the final meson
in the center-of-mass frame of the charmed meson pair. The twice logarithmic
scale of the picture reveals, that the cross section depends on \( \Delta  \)
near the threshold as: \[
\sigma _{VV},\, \sigma _{PP}\sim \Delta ^{1/2}\sim |k|,\]

\[
\sigma _{PV}\sim \Delta ^{3/2}\sim |k|^{3}.\]

The point is that in the \( \gamma \gamma  \)-interaction the \( D\bar{D} \)-
and \( D^{*}\bar{D}^{*} \)-pairs are produced in S-wave state (L=0),
and \( D\bar{D}^{*} \)-pairs are produced in P-wave state (L=1).
It is known, that near the threshold the behavior of reactions with
two particles in the final state submits to the law:

\begin{equation}
\label{sigmaotL}
\sigma \sim |k|^{2L+1}.
\end{equation}

The diagrams of Fig.\ref{diagrams} which describes the processes
\( \gamma \gamma \rightarrow D\bar{D} \), \( \gamma \gamma \rightarrow D^{*}\bar{D}^{*} \),
\( \gamma \gamma \rightarrow D\bar{D}^{*} \) are written at the quark
level and the \char`\"{}twisting\char`\"{} of the final mesons into
the necessary wave occurs automatically. Hence, the observable behavior
of the cross sections near threshold is the additional confirmation
of the calculation correctness. 

It is apparent from Fig. \ref{DDcharged}, that the HQET ratio (\ref{ratio3})
is broken. Our calculations show, that near the threshold 

\[
\sigma _{PP}:\sigma _{PV}:\sigma _{VV}\approx 1:0:13.4.\]

It is interesting to consider the case when the interaction of the
light quark with the initial photon can be neglected i.e. when the
electric charge of the light quark is put equal to zero. The black
markers in Fig. \ref{DDcharged} show the results of the calculations
of the charmed meson pair production cross sections at a zero light
quark charge. One can see, that near the threshold the ratio \( \sigma _{PP}:\sigma _{PV}:\sigma _{VV} \)
is practically the same as predicted by (\ref{ratio3}). It is interesting
to mention, that the contribution from the interaction between the
photon and the light quark can be positive (the case of \( D^{*}\bar{D}^{*} \)-pair
production), as well as negative (the case of \( D\bar{D} \) or \( D^{*}\bar{D} \)
-pair production). Let us note also, that contribution of this interaction
has the minimal absolute value for the \( D^{*}\bar{D} \) -pair production,
i.e. when the mesons are formed in the \( P \)-wave state.

\begin{figure}
{\centering \hfill\parbox{12.5cm}{\( \sigma  \), nb}\par}

{\centering \resizebox*{0.5\textwidth}{!}{\includegraphics{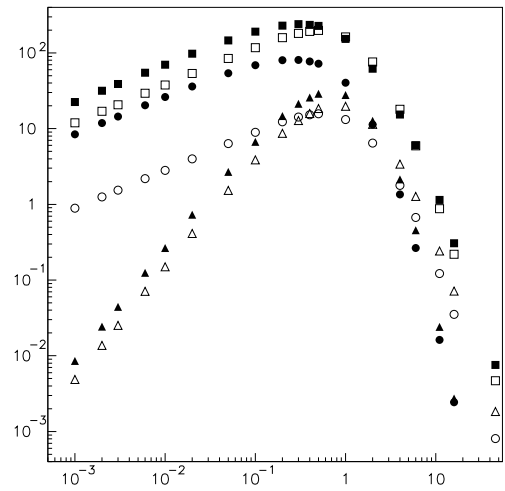}} \par}

{\centering \parbox{8cm}{\hfill \( \Delta  \), GeV}\par}

\caption{The exclusive cross section of the photonic pair production for the
charged charmed mesons \protect\( \sigma _{PP}\protect \) (open circles),
\protect\( \sigma _{PV}\protect \) (open triangles), \protect\( \sigma _{VV}\protect \)
(open squares) as function of \protect\( \Delta \protect \) in comparison
with the same dependences for the neutral mesons (black markers)\label{DD0}.}
\end{figure}

The extremely essential role of the light quark in the charmed meson
production can be understood from the difference between the production
of neutral mesons and the charged ones. This difference can be clearly
seen in Fig. \ref{DD0}, where the \( \Delta  \)-dependence of the
neutral meson cross section (the black markers) is shown in comparison
with the same dependence for charged mesons (the open ones). For example,
near the threshold for the charged mesons \( \sigma _{D^{+}D^{-}}:\sigma _{D^{+*}D^{-*}}\approx 1:13.4 \)
in the contrary to \( \sigma _{D^{0}\bar{D}^{0}}:\sigma _{D^{0*}\bar{D}^{0*}}\approx 1:2.7 \)
for the neutral ones.

It is necessary to note, that the least difference between the charged
and neutral meson production is observed for the production of \( D\bar{D}^{*} \)-pairs
i.e. when the mesons are produced in the P-wave state.

\begin{figure}
{\centering \hfill\parbox{12.5cm}{\( \sigma  \), nb}\par}

{\centering \resizebox*{0.5\textwidth}{!}{\includegraphics{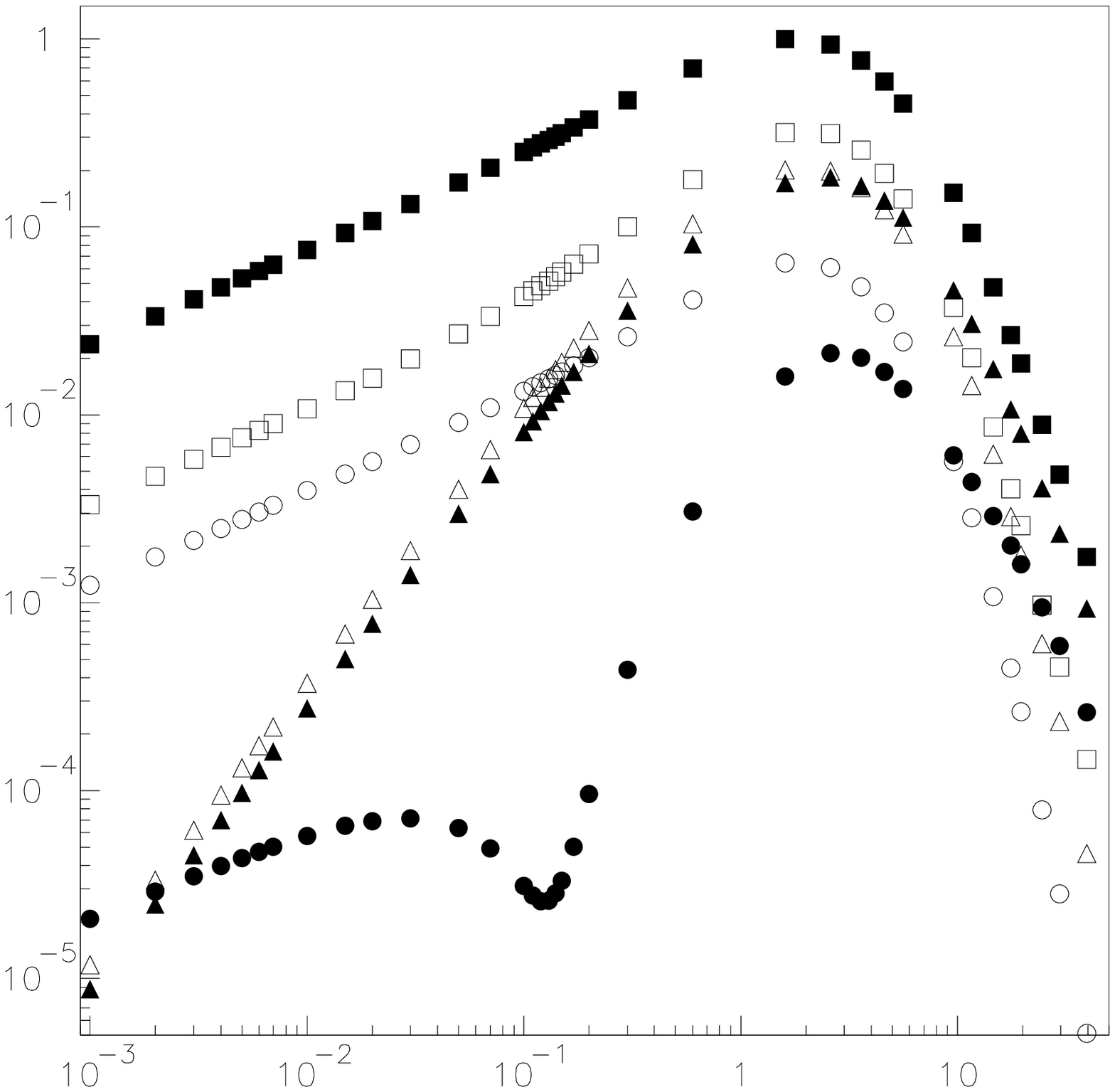}} \par}

{\centering \parbox{8cm}{\hfill \( \Delta  \), GeV}\par}

\caption{The photonic production of the charged \protect\( B\protect \)-mesons
(black markers) and the neutral ones (open markers) as a function
of \protect\( \Delta \protect \).\label{BB}}
\end{figure}

The more striking differences can be observed between the neutral
and charged B-meson production. In Fig. \ref{BB} the production cross
section of the neutral B-meson is given as a function of \( \Delta ={\sqrt{s}}-{\sqrt{s_{th}}} \)
for \( m_{b}=5 \) Gev and \( m_{q}=0.2 \). In the case of D-meson
production the cross section dependences on \( \Delta  \) have the
qualitatively identical form both for the charged particles and for
the neutral ones (see Fig. \ref{DD0}). For the B-meson production
it is not so. One can see in Fig. \ref{BB}, that at \( \Delta \approx 0.12 \)
GeV (\( |k|\approx 0.8 \) GeV) the interferential minimum is observed
for the production of two charged pseudo-scalar mesons \( B^{+} \)
and \( B^{-} \) and not observed for the pair production of the neutral
pseudo-scalar mesons \( B^{0} \) and \( \bar{B}^{0} \).

So, for all interaction energies it is necessary to take into account
the light quark contribution into the photonic production of heavy
mesons. The models which are not taking into account this contribution
can not be applied even for the qualitative description of the considered
process. It is clear from the analysis of the photonic pair production
of the heavy mesons, that the naive ideas which transferred from the
\( e^{+}e^{-} \)-annihilation appear not valid. For the processes
\( \gamma \gamma \rightarrow D\bar{D} \) and \( \gamma \gamma \rightarrow B\bar{B} \)
the role of light quarks appears essential both in the soft process
of the hadronization and in the hard pair production of the heavy
quarks. Unfortunately, the experimental data on the heavy meson pair
production near threshold do not exist yet. Such data could throw
light on the dynamics of production of mesons with the heavy quarks.

\section{The inclusive heavy meson production in the photon-photon interaction}

As it was already mentioned, the exclusive heavy meson production
data would allow to compare the predictions of various models, however
the opportunities of experiment in this direction of researches are
strongly limited. For today the majority of measurements are made
for the inclusive \( D \)-meson production. To describe this process
a number of the calculations is carried out in the framework of NLO
QCD for the massive c-quarks \cite{Frixione}, as well as for c-quarks
\cite{Kniehl}. These calculations result in the satisfactory description
of the data, however, the use of the fragmentation mechanism result
in specific predictions for the relative yield of charged and neutral
mesons. In these models the \( D^{*} \)-yield does not depend on
the meson type: \( \sigma _{D^{*0}}=\sigma _{D^{*+}} \).

Let us consider now the prediction of our model for the inclusive
heavy meson production cross section. The diagrams we need are analogous
to the ones of Fig. \ref{diagrams}. The only difference is that the
second pair of heavy and light quarks is not fused in the meson. Thus
we take into account the resonant part of the spectrum for one quark
pair and the continuum for another. To calculate the inclusive production
cross section we use the values of parameters which describes the
photoproduction of the \( D^{*} \)-mesons on the accelerator HERA
\cite{BKL}, as well as in our calculation of the exclusive charmed
meson production. To estimate the inclusive production cross section
we shall take into account the octet contribution which is absent
in the exclusive production. In the case of the photonic inclusive
production of \( D \)-mesons the octet contribution differs from
the singlet one only in the general coefficient. If one use the parameters,
from \cite{BKL}, then the octet contribution amounts 16\% of the
total cross section of the \( D \)-meson production.

\begin{figure}
\hfill\parbox{12.5cm}{\( \sigma  \), nb}

{\centering \resizebox*{0.5\textwidth}{!}{\includegraphics{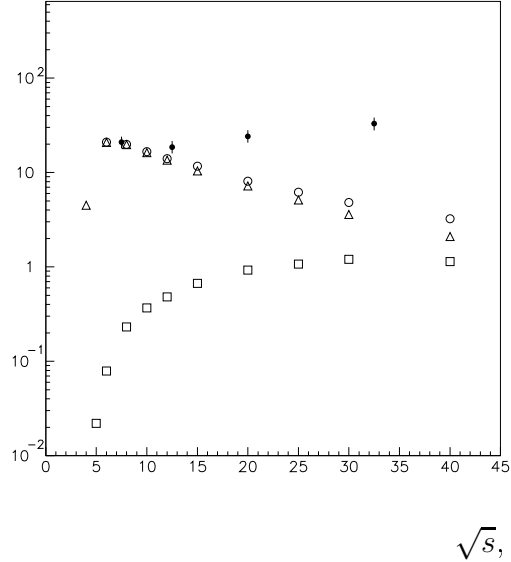}} \par}

{\centering \parbox{8cm}{\hfill \( \sqrt{s} \), GeV}\par}

\caption{The charm production cross section \protect\( \gamma \gamma \rightarrow c\bar{c}+X\protect \)
(open circles) in comparison with the L3 Collaboration experimental
data\cite{L3}. \label{sigma_s}}
\end{figure}

One can see from Fig. \ref{sigma_s} that our predictions for the
inclusive production describe L3 experiment data in the order of magnitude
(open circles). As our calculations show, the nonzero light quark
charge results in the value of ratio \( \sigma _{D^{*+}}/\sigma _{D^{*0}} \)
differ from unit. This is valid for the inclusive production too:
our calculations show, that the inclusive \( D^{*0} \)-meson yield
approximately twice surpasses the yield of the \( D^{*+} \)-meson,
while the fragmentation model predicts \( \sigma _{D^{*+}}/\sigma _{D^{*0}}=1 \). 

At energies larger than 15 GeV (the range where \( c\bar{q} \)-pair
produced mainly at large invariant masses) the total cross section
calculated in our model begins to decrease and underestimate the experimental
data. In this energy range it is not enough to consider the resonant
contribution (open triangles), and it is necessary to take into account
the contribution of continuum (open squares). To calculate the contribution
of the continuum we do not fuse now the light quarks with the \( c \)-quarks:
\( \gamma \gamma \rightarrow c\bar{c}q\bar{q} \).  In the conception
of structure functions the considered mechanism means the account
of the hadronic structures of the photon. Our approach allow to take
into account only a part of the photon structure. That is why  we
underestimate the production cross section, and the additional part
of the photon structure function is necessary to total value of the
cross section. We expect for the contribution of continuum, that the
yields of the charged mesons and neutral ones should be identical.
Thus the ratio \( \sigma _{D^{*+}}/\sigma _{D^{*0}} \) could serve
as a good indication of the production mechanism. The calculations
which have been carried out in the works \cite{Frixione,Kniehl} show
that the use of the total structure function allows to describe experimental
data well. 

\section*{Conclusion}

In the framework of CQM we have analysed the exclusive production
of the heavy mesons in the \( \gamma \gamma  \)-collisions near the
threshold, as well as the inclusive heavy meson production at high
energies. Near the threshold we have calculated the relative contribution
of the different meson pairs and have found out that the calculation
results essentially depends on the electric charge of the light constituent
quark. For zero electric charge of the light quark we obtain, that
the production process looks like the heavy quark production followed
hadronization according to HQET. At the real values of charges the
strong deviation from the predictions of the HQET takes place. Thus,
our analysis of the photon-photon production of the heavy meson pairs
shows, that the naive conception which is transferred from the \( e^{+}e^{-} \)-annihilation
appear not valid. In the processes \( \gamma \gamma \rightarrow D\bar{D} \)
and \( \gamma \gamma \rightarrow B\bar{B} \) the role of the light
quarks is essential both in the soft hadronization process and in
the hard process of the heavy quark pair production. Unfortunately,
in the framework of our model it is impossible to use the same values
of constants \( f_{P} \) and \( f_{V} \) to calculate both the exclusive
channel and the inclusive one. The use of the values of \( f_{P} \)
and \( f_{V} \), which describes process \( \gamma p\rightarrow D^{*}+X \),
leads to the large overestimate of the production cross section near
the threshold. We suppose, that such overestimation takes place because
the values \( f_{P} \) and \( f_{V} \) effectively account for the
contributions from the decays of the excited states, for example,
from the decay \( D^{**}\rightarrow D^{*}+X \). So, these values
can not be applied to the case of the exclusive meson production.
Nevertheless, our model gives reasonable estimation of the relative
contributions into the cross section of the different heavy meson
pairs (\( D^{0}\bar{D}^{0} \), \( D^{0*}\bar{D}^{0} \), \( \bar{D}^{0}D^{0*} \),
\( D^{0*}\bar{D}^{0*} \), \( D^{+}D^{-} \), \( D^{+}\bar{D}^{-*} \),
\( D^{-}D^{+*} \), \( D^{+*}D^{-*} \)), as well as reasonable predictions
of the dependence of the cross section on the interaction energy and
the shape of the differential distributions. 

Also in the framework of the CQM we calculated the cross section of
the inclusive charmed meson production in the photon-photon interaction.
The cross section prediction of the inclusive production calculated
in the frame work of CQM is in accordance with the experiment data
at low energies of the \( \gamma \gamma  \)-interactions if one choose
the parameter values which describes the process \( \gamma p\rightarrow D^{*}+X \)
at HERA \cite{BKL}. It is not unexpected for us because the contribution
of the low \( \gamma g \)-interaction energies is the most essential
for the \( D^{*} \)-meson photoproduction at the accelerator HERA
. The continuum contribution should increase as a function of the
interaction energy. Thus at large energies the character of the calculations
should be the same, as in the works \cite{Frixione,Kniehl}, which
explain the cross section increasing for the process \( \gamma \gamma \rightarrow c\bar{c} \)
with the help of the account for the contribution of the light constituents
in the initial photon structure. 

The work is supported in part by the grants of RFBR 99-02-6558 and
00-15-96645, the grant of the Education Ministry of Russian Federation
E00-33-062 and the grant CRDF MO-001-0.

\section*{Appendix}

In the given appendix we show how to derive the ratio \( \sigma _{PP}:\sigma _{PV}:\sigma _{VV} \)
near the threshold for the heavy meson production in the \( e^{+}e^{-} \)-annihilation,
as well as in the \( \gamma \gamma  \) - interaction in the approach
of the heavy quark effective theory. 

Let us accept the following designations: 

\( Q_{+(-)} \) (\( \bar{Q}_{+(-)} \)) is the wave function of the
heavy quark (antiquark) with the spin projection \( +1/2(-1/2) \)
on the chosen axis; 

\( q_{+(-)} \) ( \( \bar{q}_{+(-)} \)) is the wave functions of
the light quark (antiquark). 

In the framework of HQET the heavy quarks are produced in the \( e^{+}e^{-} \)-annihilation
the by means of the virtual photon decay, so, these quarks are in
\( J^{P}=1^{-} \) state. The spin part of the wave function of such
state can be expressed through the spinor wave functions of the quarks
as follows:\[
\hat{\Psi }_{Q\bar{Q}}=(\Psi _{Q\bar{Q}}^{+1},\Psi _{Q\bar{Q}}^{0},\Psi _{Q\bar{Q}}^{-1})=(Q_{+}\bar{Q}_{+},\frac{Q_{+}\bar{Q}_{-}+Q_{-}\bar{Q}_{+}}{\sqrt{2}},Q_{-}\bar{Q}_{-}).\]

Thus the light quarks should be in the \( 0^{+} \)-state. This state
can be obtained in the P-wave as follows:\[
\psi _{q\bar{q}}=(n_{-1}\Psi _{q\bar{q}}^{+1}+n_{0}\Psi _{q\bar{q}}^{0}+n_{+1}\Psi _{q\bar{q}}^{-1})/\sqrt{3},\]
where \( \hat{n} \) is the orbital part of the wave function of the
light quark pair, and \( \hat{\Psi }_{q\bar{q}} \) is the spin part
of the one:\[
\hat{\Psi }_{q\bar{q}}=(\Psi _{q\bar{q}}^{+1},\Psi _{q\bar{q}}^{0},\Psi _{q\bar{q}}^{-1})=(q_{+}\bar{q}_{+},\frac{q_{+}\bar{q}_{-}+q_{-}\bar{q}_{+}}{\sqrt{2}},q_{-}\bar{q}_{-}).\]

So, the total wave function looks like follows:

\begin{equation}
\label{totalwf}
\hat{\Psi }=\hat{\Psi }_{Q\bar{Q}}\cdot (n_{-1}\Psi _{q\bar{q}}^{+1}+n_{0}\Psi _{q\bar{q}}^{0}+n_{+1}\Psi _{q\bar{q}}^{-1})/\sqrt{3}.
\end{equation}

The right part of equation (\ref{totalwf}) includes the products
of the spinor wave functions of the quarks \( Q_{+(-)}\bar{q}_{+(-)} \)
and \( \bar{Q}_{+(-)}q_{+(-)} \), which can be expressed through
the spin functions of the final mesons:\begin{equation}
\label{pwf}
P=\frac{Q_{+}\bar{q}_{-}-Q_{-}\bar{q}_{+}}{\sqrt{2}},
\end{equation}
where \( P \) is the wave function of the pseudoscalar meson;\begin{equation}
\label{vwfplus}
V_{+1}=Q_{+}\bar{q}_{+,}
\end{equation}

\begin{equation}
\label{vwf0}
V_{0}=\frac{Q_{+}\bar{q}_{-}+Q_{-}\bar{q}_{+}}{\sqrt{2}},
\end{equation}
\begin{equation}
\label{vwfminus}
V_{-1}=Q_{-}\bar{q}-,
\end{equation}
where \( P \) and \( V \) are the wave functions of the pseudoscalar
and vector mesons, correspondingly.

One can conclude from (\ref{pwf}) and (\ref{vwf0}) that\begin{equation}
\label{vpplus}
Q_{+}\bar{q}_{-}=\frac{V_{0}+P}{\sqrt{2}},
\end{equation}

\begin{equation}
\label{vpminus}
Q_{-}\bar{q}_{+}=\frac{V_{0}-P}{\sqrt{2}}.
\end{equation}

From the equations (\ref{vwfplus},\ref{vwfminus},\ref{vpplus},\ref{vpminus})
the following expressions for the components of the total wave function
(\ref{totalwf}) can be derived :\[
\Psi ^{+1}=\frac{1}{2\sqrt{3}}\Bigl (n_{+1}P\bar{P}+\{n_{+1}P\bar{V}_{0}+n_{+1}V_{0}\bar{P}+n_{0}P\bar{V}_{+1}+n_{0}V_{+1}\bar{P}\}+\]
\begin{equation}
\label{totalplus}
+[n_{+1}V_{0}\bar{V}_{0}+n_{0}V_{0}\bar{V}_{+1}+n_{0}V_{+1}\bar{V}_{0}+2n_{-1}V_{+1}\bar{V}_{+1}]\Bigr ),
\end{equation}
\[
\Psi ^{0}=\frac{1}{2\sqrt{3}}\Bigl (-n_{0}P\bar{P}+\{n_{+1}P\bar{V}_{-1}+n_{+1}V_{-1}\bar{P}-n_{-1}P\bar{V}_{+1}-n_{-1}V_{+1}\bar{P}\}+\]
\begin{equation}
\label{total0}
+[n_{+1}V_{0}\bar{V}_{-1}+n_{+1}V_{-1}\bar{V}_{0}+n_{0}V_{0}\bar{V}_{0}+n_{0}V_{+1}\bar{V}_{-1}+n_{0}V_{-1}\bar{V}_{+1}+n_{-1}V_{+1}\bar{V}_{0}+n_{-1}V_{0}\bar{V}_{+1}]\Bigr ),
\end{equation}
\[
\Psi ^{-1}=\frac{1}{2\sqrt{3}}\Bigl (n_{-1}P\bar{P}-\{n_{-1}P\bar{V}_{0}+n_{-1}V_{0}\bar{P}+n_{0}P\bar{V}_{-1}+n_{0}V_{-1}\bar{P}\}+\]
\begin{equation}
\label{totalminus}
+[n_{-1}V_{0}\bar{V}_{0}+n_{0}V_{0}\bar{V}_{-1}+n_{0}V_{-1}\bar{V}_{0}+2n_{+1}V_{-1}\bar{V}_{-1}]\Bigr ).
\end{equation}

As we believe, that the spins of the heavy quark and the light quark
do not interact with each other and with the orbital moment, the all
term in the right part of the equations (\ref{totalplus},\ref{total0},\ref{totalminus})
give the identical contributions to the square of the total wave function
\( \hat{\Psi } \) . The result (\ref{eeotn}) becomes obvious, because
all these terms are orthogonal to each other. 

In the case of \( \gamma \gamma  \)-interactions two initial photons
can be in the following states: \( 0^{+}, \) \( 0^{-}, \) \( 2^{+} \).
And the \( 0^{+} \)-state can be produced in the \( S \)-wave only.
Since at the threshold survive contributions of the state with a minimal
wave number (see (\ref{sigmaotL})), we shall consider \( 0^{+} \).
Such state can be made of two pseudo-scalar functions:

\[
\Psi _{Q\bar{Q}}=\frac{Q_{+}\bar{Q}_{-}-Q_{-}\bar{Q}_{+}}{\sqrt{2}}\]

and\[
\Psi _{q\bar{q}}=\frac{q_{+}\bar{q}_{-}-q_{-}\bar{q}_{+}}{\sqrt{2}}\]
by the following manner:

\begin{equation}
\label{totalwf2ph}
\Psi =\Psi _{Q\bar{Q}}\cdot \Psi _{q\bar{q}}.
\end{equation}

Using (\ref{vwfplus},\ref{vwfminus},\ref{vpplus},\ref{vpminus})
to transform the equation (\ref{totalwf2ph}), we obtain:

\[
\Psi =\frac{1}{2}(-P\bar{P}+V_{0}\bar{V}_{0}-V_{-1}\bar{V}_{+1}-V_{+1}\bar{V}_{-1}),\]
whence the ratio for the photonic production near the threshold follows:\[
\sigma _{PP}:\sigma _{PV}:\sigma _{VV}=1:0:3\]

\end{document}